\documentclass[apjl,aas_macros]{emulateapj}

\usepackage{apjfonts}
\usepackage{color}
\usepackage{amsmath}
\usepackage{textcomp}
\usepackage{url}
\usepackage{graphicx, subfigure}

\newcommand{\kms}{\hbox{km~s$^{-1}$}}
\newcommand{\Mjup}{$M_{\mathrm{Jup}}$}
\newcommand{\masyr}{$\mathrm{mas}\ \mathrm{yr}^{-1}$}

\slugcomment{To appear in ApJ Letters.}

\shorttitle{A New L4$\beta$ Candidate Member of Argus}
\shortauthors{Gagn\'e et al.}

\begin{document}

\title{SIMP~J2154--1055: A NEW LOW-GRAVITY L4$\beta$ BROWN DWARF CANDIDATE MEMBER OF THE ARGUS ASSOCIATION}

\author{Jonathan Gagn\'e\altaffilmark{1},\, David Lafreni\`ere\altaffilmark{1},\, Ren\'e Doyon\altaffilmark{1},\, \'Etienne Artigau\altaffilmark{1},\, Lison Malo\altaffilmark{1,2},\, Jasmin Robert\altaffilmark{1} and Daniel Nadeau\altaffilmark{1}}
\affil{\altaffilmark{1} D\'epartement de Physique, Universit\'e de Montr\'eal, C.P. 6128 Succ. Centre-ville, Montr\'eal, QC H3C 3J7, Canada}
\affil{\altaffilmark{2} Canada-France-Hawaii Telescope, 65-1238 Mamalahoa Hwy, Kamuela, HI 96743, USA}

\begin{abstract}
We present SIMP~J21543454--1055308, a new L4$\beta$ brown dwarf identified in the \emph{Sondage Infrarouge de Mouvement Propre} survey that displays signs of low gravity in its near-infrared spectrum. Using the Bayesian Analysis for Nearby Young AssociatioNs II (BANYAN~II), we show that it is a candidate member of the Argus association, albeit with a 21\% probability that it is a contaminant from the field. Measurements of radial velocity and parallax will be needed to verify its membership. If it is a member of Argus (age 30--50~Myr), then this object would have a planetary mass of $10 \pm 0.5$ \Mjup.
\end{abstract}

\keywords{brown dwarfs --- methods: data analysis --- proper motions --- stars: kinematics and dynamics}

\section{INTRODUCTION}

In the last decade, several brown dwarfs (BDs) in the field have been identified as displaying low-gravity features attributable to youth (\citealp{2008ApJ...689.1295K}; \citealp{2009AJ....137.3345C}; \citealp{2013ApJ...772...79A}; \citealp{2013ApJ...777L..20L}; \citealp{2014ApJ...785L..14G}). One could expect that a fraction of those young objects are unrecognized members of kinematic associations. However, the lack of parallax or radial velocity measurements for those objects prevents us from directly assessing their kinematics, which makes the identification of BD members to nearby young associations very hard. Identifying such BDs of known age would provide benchmarks to study the atmospheres of very low-mass BDs that are known to have features similar to those of the few directly imaged giant exoplanets known today (e.g., \citealp{2004A&A...425L..29C}; \citealp{2008Sci...322.1348M}; \citealp{2008ApJ...689L.153L}; \citealp{2009A&A...493L..21L};  \citealp{2013ApJ...774...55B}; \citealp{2013ApJ...779L..26R}; \citealp{2013ApJ...774...11K}; \citealp{2014ApJ...780L..30C}; \citealp{2014ApJ...787....5N}). In order to address this, \cite{2014ApJ...783..121G} developed the Bayesian Analysis for Nearby Young AssociatioNs II tool (BANYAN~II), a statistical analysis based on BANYAN~I \citep{2013ApJ...762...88M} that uses a naive Bayesian classifier to identify low-mass star and BD candidates to membership in nearby young associations, even in the absence of radial velocity and parallax measurements. \\

SIMP~J21543454--1055308 (SIMP~J2154--1055 hereafter) is a new low-gravity BD that we discovered as part of the \emph{Sondage Infrarouge de Mouvement Propre} (SIMP). We briefly describe the SIMP survey in Section~\ref{sec:simp}, and then present a spectroscopic follow-up of SIMP~J2154--1055 in Section~\ref{sec:spectro}. In Section~\ref{sec:lowg}, we show evidence that this BD displays signs of low gravity, by comparing its near-infrared (NIR) spectrum with low-gravity standards and by using the gravity classification scheme of \cite{2013ApJ...772...79A}. We finally use the BANYAN~II tool to show that this young BD is a candidate Argus association member (Section~\ref{sec:NYAs}).

\section{THE SIMP SURVEY}\label{sec:simp}

The SIMP survey \citep{2009AIPC.1094..493A} has been initiated in 2006 to identify new nearby BDs from their red optical-to-NIR colors and high proper motions, by obtaining a second $J$-band epoch for 30\% of the sky (mostly in the Southern Hemisphere), 5--8 yr after the Two Micron All Sky Survey (2MASS; \citealp{2006AJ....131.1163S}). This was achieved by using the CPAPIR camera \citep{2004SPIE.5492.1479A} on both the CTIO~1.5m telescope and the Observatoire du Mont-M\'egantic 1.6m telescope. The SIMP survey typically reached an astrometric precision of 0\farcs15 and a photometric depth of $J=17$. Any source in the SIMP survey that satisfied at least one of three filters based on proper motion ($\mu$) and $I-J$ color was selected for a spectroscopic follow-up: (1) $\mu > 100$ \masyr\ and detected in $I$ with $I-J > 3.5$; (2) $\mu > 100$ \masyr\ and not detected in $I$ such that $I-J > 3$; or (3) $\mu > 200$ \masyr\ and not detected in $I$. $I$ magnitudes were selected either from the \emph{SuperCOSMOS Sky Survey} \citep{2001MNRAS.326.1279H}, the Sloan Digital Sky Survey \citep{2000AJ....120.1579Y}, or the Catalina Sky Survey (CSS;  \citealp{1998BAAS...30.1037L}), as available and in this order of preference. These criteria were designed to reject close-by M-type dwarfs as well as distant objects. A large number of new BD candidates have been identified this way, from which more than a hundred have been followed with NIR spectroscopy and confirmed as new M5--T3 very low-mass stars and BDs, most of them having spectral types later than L0. Among those, we have identified SIMP~J2154--1055, an L4$\beta$ BD displaying signs of low-gravity. A few other SIMP discoveries have been highlighted in \cite{2006ApJ...651L..57A} and \cite{2011ApJ...739...48A}, whereas remaining discoveries will be presented in an upcoming paper (J. Robert et al., in preparation).\\

\section{SPECTROSCOPIC FOLLOW-UP}\label{sec:spectro}

The NIR spectrum of SIMP~J2154--1055 presented here was obtained on 2008 September 16 at the IRTF (program number 2008B054), using SpeX in the prism mode with the 0\farcs5 slit ($R \sim$ 150), covering the 0.8--2.5 $\mu$m range. The source was moved along the slit in an ABBA pattern, with a total of 10 exposures of 180 s to achieve a signal-to-noise ratio of $\sim$180 per resolution element. Raw exposures were reduced and combined using the SpeXtool package \citep{2004PASP..116..362C}, and telluric corrections were applied in a standard way \citep{2003PASP..115..389V}, using the A-type star HD~202990 observed immediately before the target and at a similar airmass.

\begin{figure*}
	\centering
	\label{fig:NIR_compb}\includegraphics[width=0.99\textwidth]{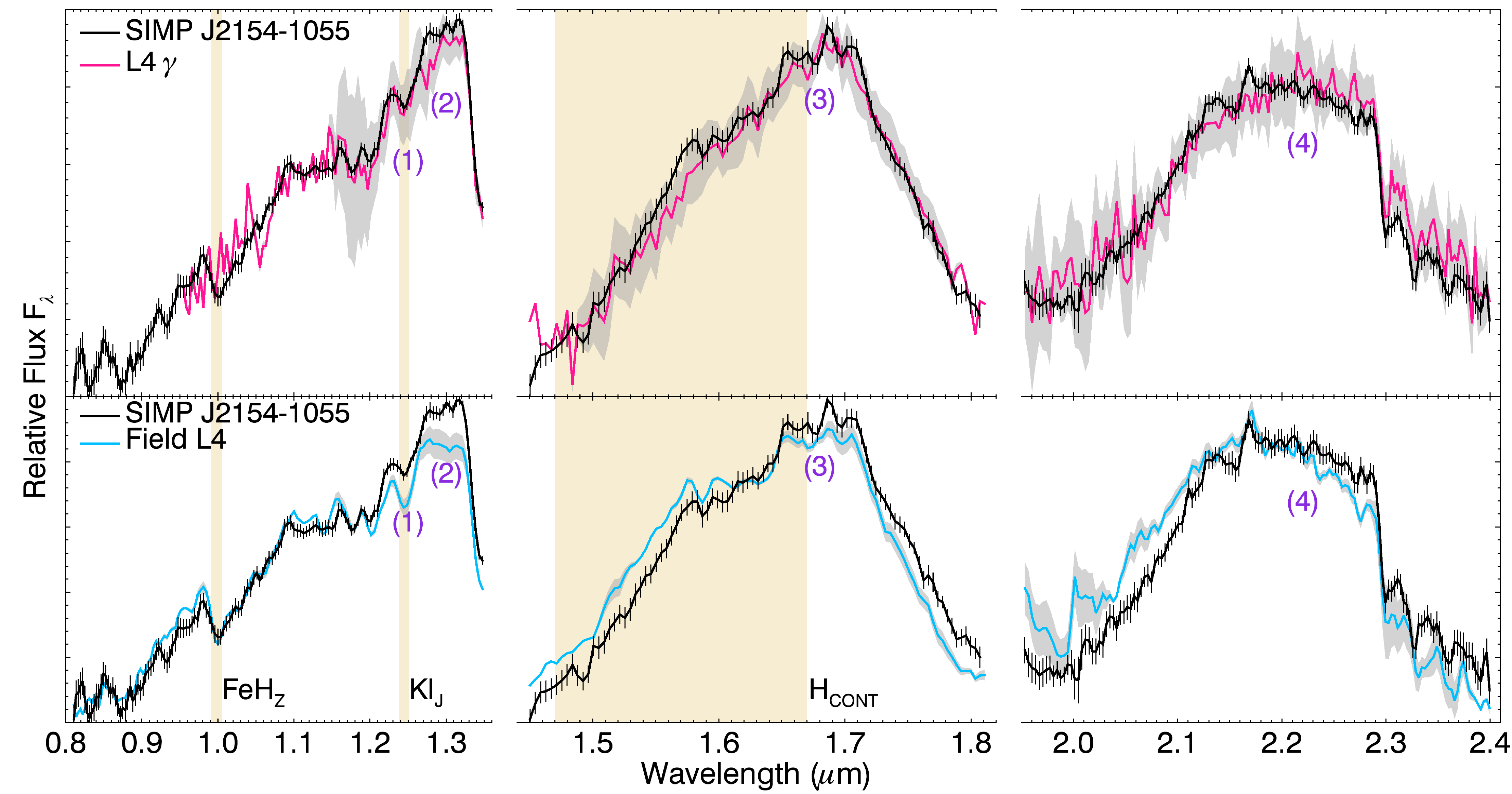}
	\caption{NIR spectrum of SIMP~J2154-1055, compared with field (blue) and very low-gravity (red) template spectra of the same spectral type. Each band was normalized individually. We classify this object as L4$\beta$ using the classification scheme of \citeauthor{2013ApJ...772...79A} (\citeyear{2013ApJ...772...79A}; Section~\ref{sec:lowg}), however we do not currently have access to any L4$\beta$ BD to build a template and compare it with SIMP~J2154--1055. The gray shaded region represents the scatter of the individual objects used to create the templates and the black vertical lines represent the measurement uncertainty on each bin of the observed spectrum. Beige shaded regions correspond to the locations of gravity-sensitive spectral indices defined by \cite{2013ApJ...772...79A}. The global continuum shape of this new BD is a better match to the L4$\gamma$ template, despite its \emph{H}-band continuum that seems to be an intermediate case between the two templates, which is consistent with our classification. We denote regions useful to differentiate between the field and low-gravity templates with purple numbers (Section~\ref{sec:lowg}).}
	\label{fig:SPT_seq}
\end{figure*}

\section{SIGNS OF LOW-GRAVITY}\label{sec:lowg}

The NIR spectrum of SIMP~J2154--1055 was visually compared with various NIR spectral templates to assign it a spectral type and identify peculiar features, following the method of K. Cruz et al. (in preparation;  see also \citealp{DisentanglingLDwar:db}). Field templates were built for all spectral types in the M5--T8 range by median-combining high-quality spectra that showed no peculiarities nor signs of unresolved binarity in the Dwarf\-Archives.\footnote[1]{\url{http://dwarfarchives.org}} Low-gravity templates were built from known L0$\gamma$--L4$\gamma$ BDs that were classified as low-gravity objects in the optical or the NIR (\citealp{2009AJ....137.3345C}; \citealp{2013ApJ...772...79A}; K. Cruz et al., in preparation). This comparison showed that SIMP~J2154--1055 matches the L4 templates better than any other spectral types. Furthermore, the L4$\gamma$ template is a better match than the field L4. The key spectral regions that differentiate between those two templates are; (1) the depth of the $KI_J$ feature at $\sim$1.25~$\mu$m; (2) the level of the red end of the $J$ band ($\sim$1.3~$\mu$m); (3) the triangular shape of the $H$ band (1.5--1.8~$\mu$m); and (4) the slope of the $K$ band plateau (2.15--2.3~$\mu$m). We show a comparison of SIMP~J2154--1055 with the L4 and L4$\gamma$ templates in Figure~\ref{fig:SPT_seq}, with those four regions identified. The L4 and L4$\gamma$ templates were built using five and two distinct spectra respectively.\\

\begin{figure*}
	\centering
	\subfigure[FeH$_Z$]{\label{fig:FeH_Z}\includegraphics[width=0.49\textwidth]{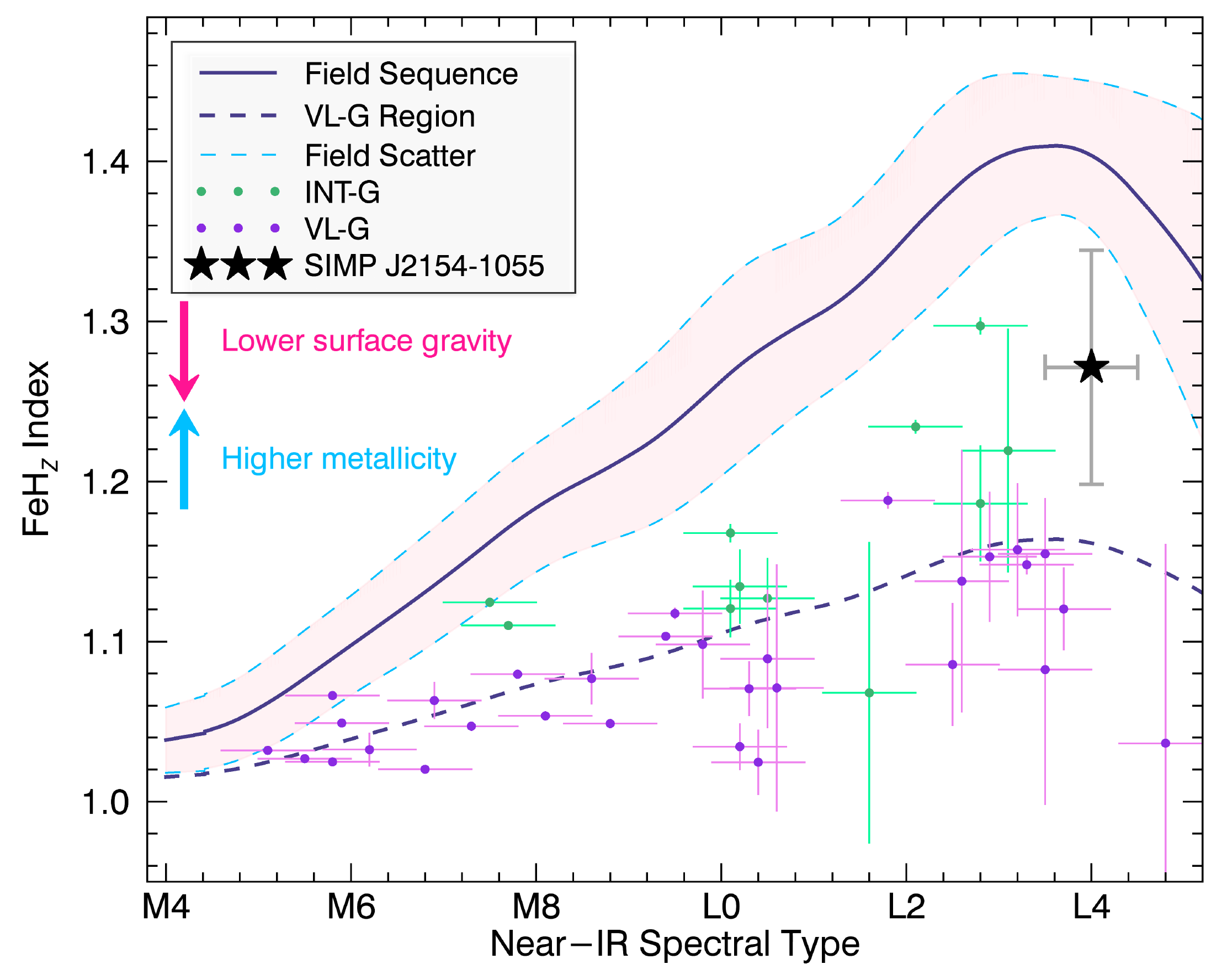}}
	\subfigure[H$_{\mathrm{CONT}}$]{\label{fig:H_CONT}\includegraphics[width=0.49\textwidth]{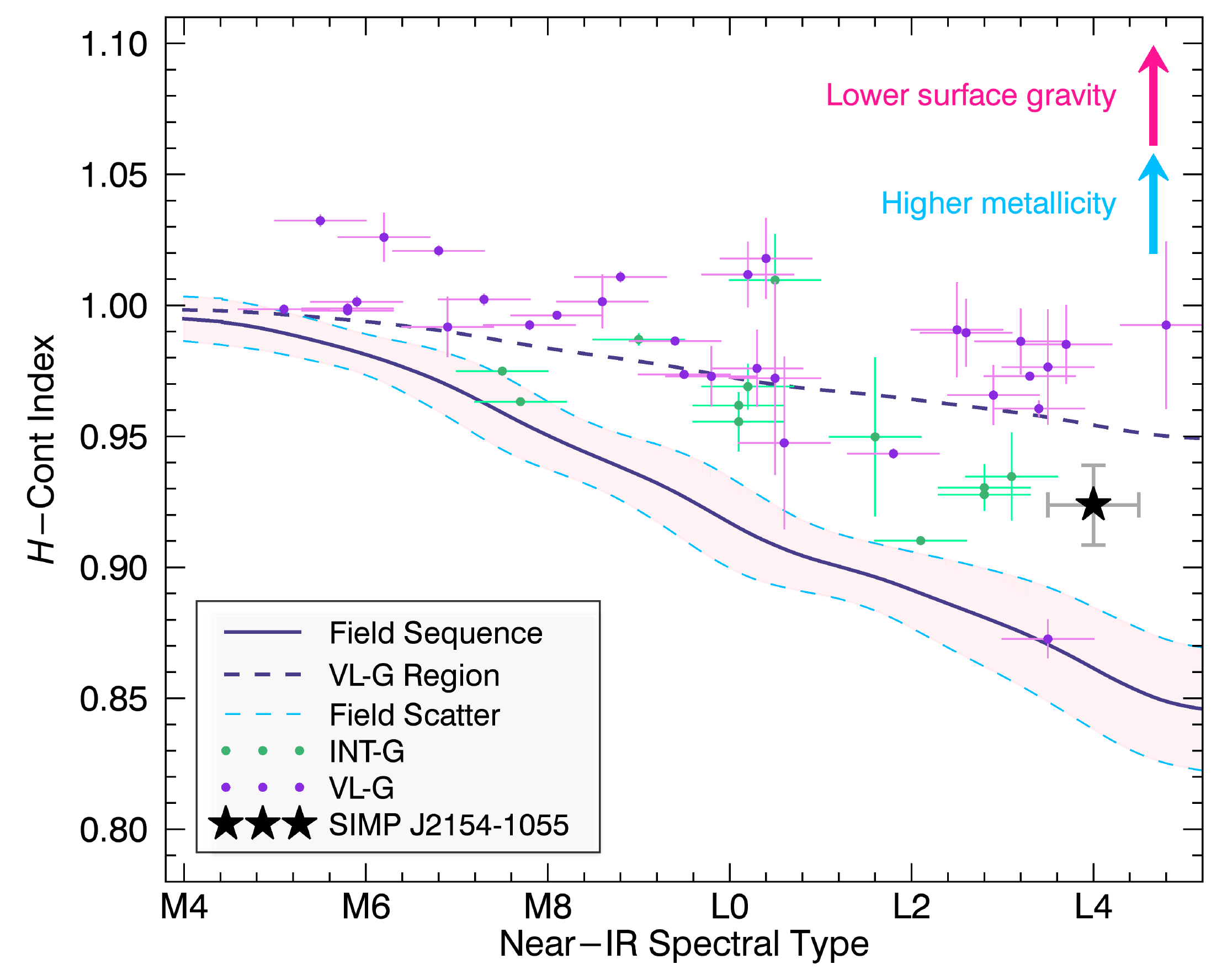}}
	\subfigure[KI$_J$]{\label{fig:KI_J}\includegraphics[width=0.49\textwidth]{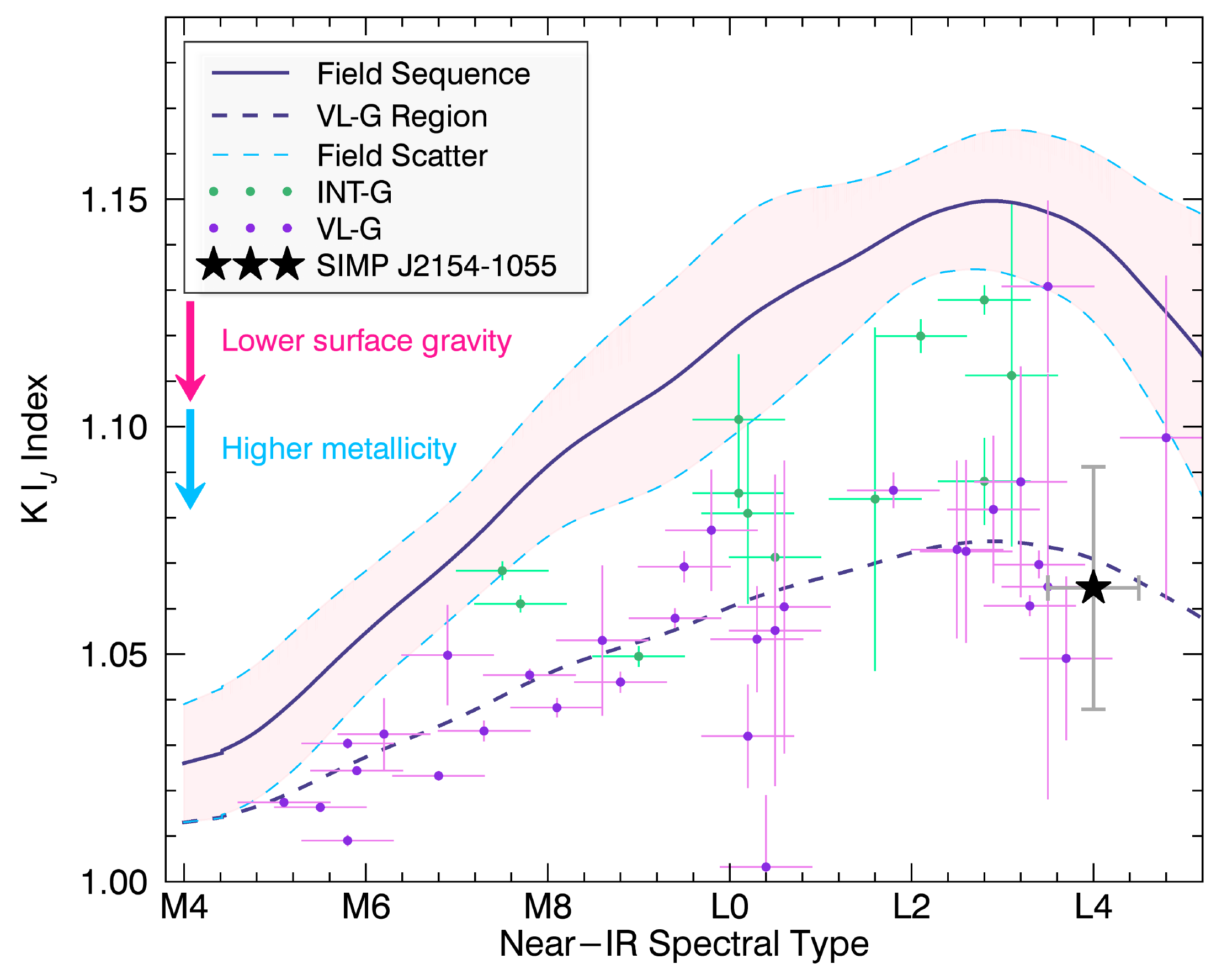}}
	\subfigure[$J-K_S$]{\label{fig:J_K}\includegraphics[width=0.49\textwidth]{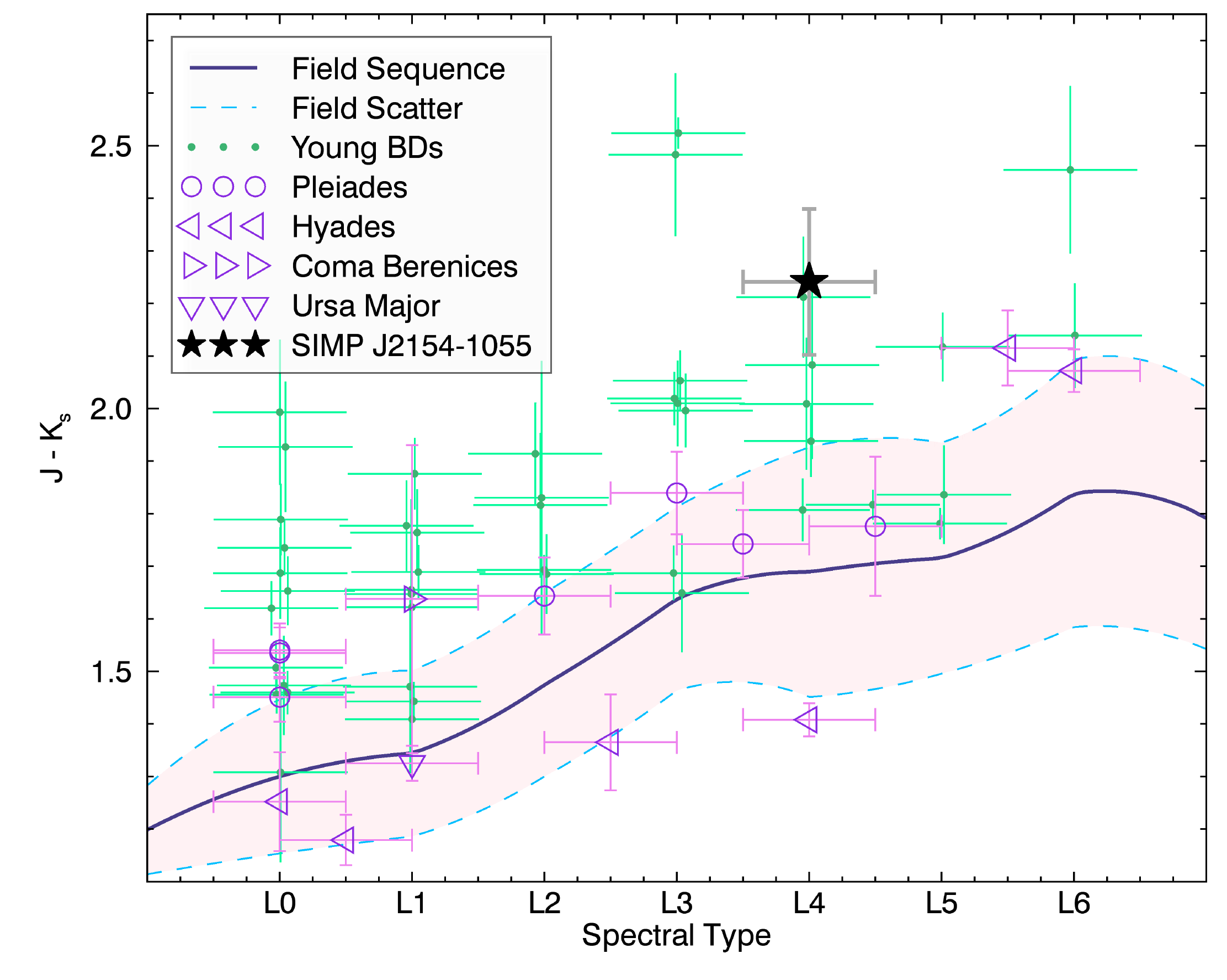}}
	\caption{Panels (a)-(c): Gravity-sensitive spectroscopic indices for SIMP~J2154--1055 (black star) compared with the field dwarf sequence (dark blue line) and its associated scatter (pink shaded region delimited by dashed pale blue lines). The dashed dark blue line delimits the intermediate-gravity and very low-gravity regimes. All objects classified as INT-G and VL-G in \cite{2013ApJ...772...79A} are also displayed as green and purple dots, respectively.\\
Panel (d): 2MASS $J-K_S$ colors of SIMP~J2154--1055 (black star) compared with the median values and standard deviation for field dwarfs (pink shaded region delimited by dashed pale blue lines; constructed from field BDs in the DwarfArchives), young field BDs (red dots; see \citealp{2014ApJ...783..121G} for the extensive list), and moderately young BD candidates (open purple symbols; \citealp{2005AN....326.1020B}; \citealp{2008MNRAS.385.1771J}; \citealp{2010A&A...519A..93B}; \citealp{2014MNRAS.441.2644C}); circles for the Pleiades ($\sim$125 Myr), downside triangles for Ursa Major ($\sim$400 Myr), right triangles for Coma Berenices ($\sim$500 Myr) and left triangles for the Hyades ($\sim$625 Myr). The colors of SIMP~J2154--1055 are consistent with its probable membership in Argus (Section~\ref{sec:NYAs}).}
\label{fig:Allers}
\end{figure*}

\begin{figure}
	\begin{center}
 	\includegraphics[width=0.495\textwidth]{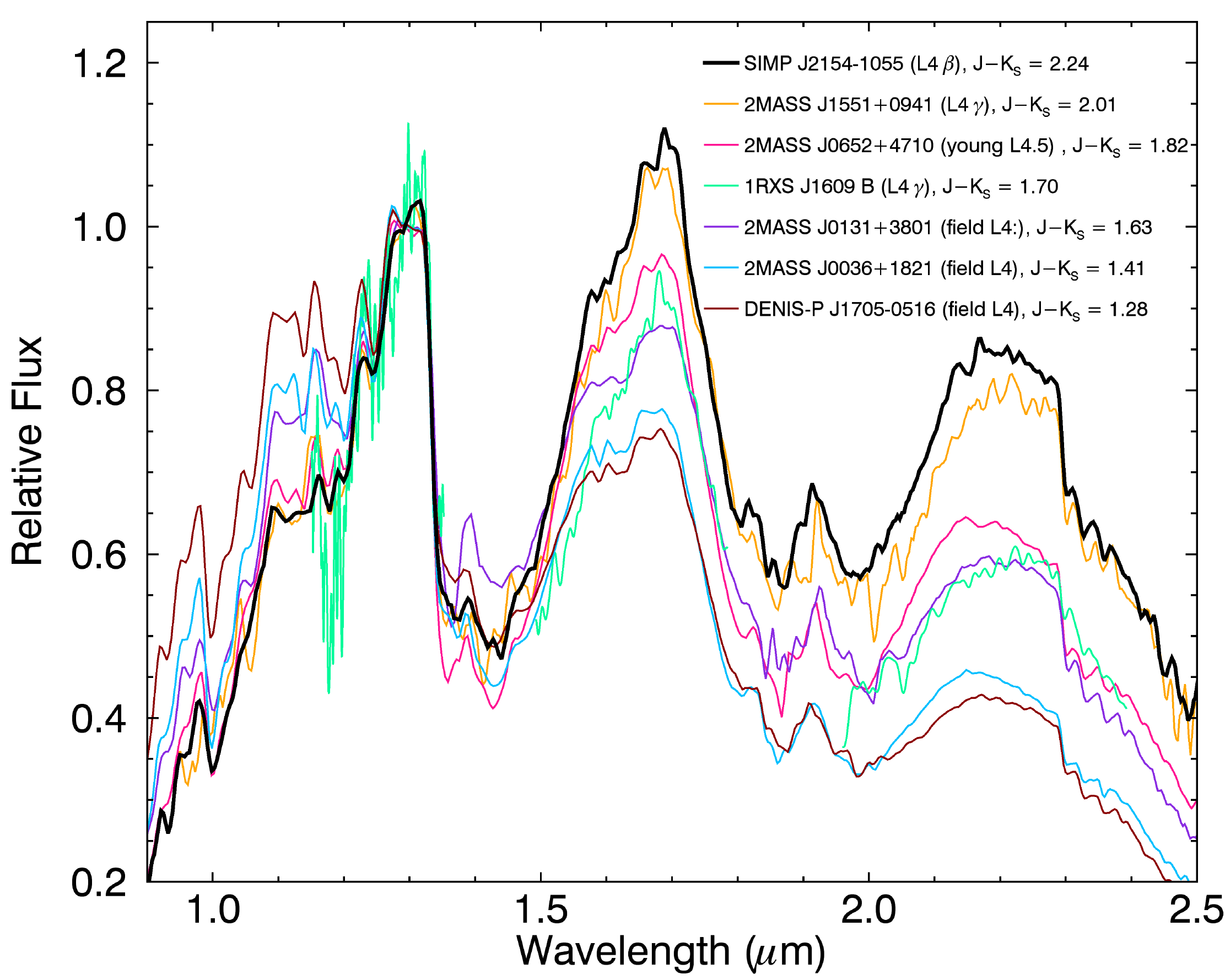}
	\end{center}
	\caption{NIR spectrum of SIMP~J2154--1055 (L4$\beta$) compared with other known L4 BDs (\citealp{2000AJ....119..369R}; \citealp{2003AJ....126.2421C}; \citealp{2004A&A...416L..17K}; \citealp{2007AJ....133..439C}; \citealp{2008AJ....136.1290R}; \citealp{2008ApJ...681..579B}; \citealp{2008ApJ...689L.153L}; \citealp{2010ApJ...719..497L}; \citealp{2010ApJ...710.1142B}; \citealp{2013ApJ...772...79A}). Although it is classified as INT-G, it is the reddest L4 BD yet identified. All spectra are normalized to their median in the 1.27--1.33 $\mu$m range.}
\label{fig:2154}
\end{figure}

\cite{2013ApJ...772...79A} developed an NIR classification scheme to determine spectral types in a way that should not be sensitive to surface gravity, using visual classification supplemented by the H$_2$O indices, and then determined a gravity class using various spectroscopic indices sensitive to surface gravity. Objects for which most low-gravity indices are strong are classified as very low-gravity (VL-G) objects, and those having only a few indices as intermediate-gravity (INT-G) objects. Those without significant signs of youth are classified as field gravity. We have used this scheme to classify SIMP~J2154--1055, with the exception that we used solely visual classification to determine its spectral type. Results are presented in Table~\ref{tab:mass}, as well as in Figure~\ref{fig:Allers}. We find it is classified as an INT-G BD. \cite{2013ApJ...772...79A} indicate that their INT-G and VL-G classifications respectively correspond to the $\beta$ and $\gamma$ classifications defined for optical spectra by \cite{2009AJ....137.3345C}. For this reason, we adopt L4$\beta$ as the NIR spectral type of SIMP~J2154--1055. Obtaining a higher-resolution NIR spectrum for this object would be useful in verifying that its alkali line equivalent widths are effectively weaker than normal.
SIMP~J2154--1055 is detected in the ALLWISE \citep{2014ApJ...783..122K} $W3$ channel, but does not display signs of infrared excess. We also show that it displays the reddest $J-K_s$ NIR color of all currently known L4 dwarfs (Figure~\ref{fig:2154}), an effect likely attributed to thicker clouds in its photosphere. It is suspected that low gravity (youth) can cause such thicker clouds \citep{2008ApJ...674..451B}.

% Table of properties
\begin{deluxetable}{lc}
\tablecolumns{2}
\tablecaption{Properties of SIMP~J2154--1055 \label{tab:mass}}
\tablehead{\colhead{Property} & \colhead{Value}}
\startdata
R.A. & 21:54:34.54\\
Decl. & --10:55:30.8\\
$\mu_\alpha\cos\delta$ (mas yr$^{-1}$)\tablenotemark{a} & $169.2\pm 8.6$\\
$\mu_\delta$ (mas yr$^{-1}$)\tablenotemark{a} & $-1.6\pm 8.8$\\
$I$ (CSS)\tablenotemark{b} & $21.63 \pm 0.06$\\
$J$ (2MASS) & $16.44 \pm 0.12$\\
$H$ (2MASS) & $15.07 \pm 0.08$\\
$K_s$ (2MASS) & $14.20 \pm 0.07$\\
$W1$ (ALLWISE) & $13.36 \pm 0.03$\\
$W2$ (ALLWISE) & $12.91 \pm 0.03$\\
$W3$ (ALLWISE) & $12.54 \pm 0.53$\\
$W4$ (ALLWISE) & $>~9.05$\\
NIR spectral type & L4 $\beta \pm 0.5$\\
Gravity score\tablenotemark{c} & \emph{1021}\\
Gravity class\tablenotemark{c} & INT-G\\
Estimated mass (\Mjup) & 10--11
\enddata
\tablenotetext{a}{Proper motion from 2MASS, ALLWISE, SIMP and DENIS.}
\tablenotetext{b}{Combined measurement from 34 epochs.}
\tablenotetext{c}{See \cite{2013ApJ...772...79A}. Scores of \emph{0}, \emph{1} or \emph{2} mean high, medium and low gravity, respectively. The four numbers are based on \emph{FeH}, \emph{VO}, \emph{K I} and the $H$-band continu\-um shape, respectively.}
\end{deluxetable}

\section{ARGUS MEMBERSHIP}\label{sec:NYAs}

We used the BANYAN~II tool \citep{2014ApJ...783..121G} to verify if SIMP~J2154-1055 is a candidate member of nearby young associations. We used the 2MASS and ALLWISE astrometry, as well as data from the DEep Near Infrared Survey of the Southern Sky (DENIS; \citealp{1998IAUS..179..106E}) and measurements obtained with CPAPIR through the SIMP survey to calculate its proper motion (Table~\ref{tab:mass}). We find that SIMP~J2154--1055 has a 83.8\% probability of being a member of the Argus association (30--50~Myr; \citealp{2008hsf2.book..757T}). The Argus association of stars was first identified by \cite{2000MNRAS.317..289M}, which proposed that the open cluster IC~2391 was a part of it. \cite{2003ASSL..299...83T} used the method of convergent point proper motion to show that both associations shared common kinematics, thus confirming this hypothesis. This association currently has 11 known A0--M5 bona fide members \citep{2013ApJ...762...88M}, from which the latest-type is the nearby star AP~Col \citep{2011AJ....142..104R}. \cite{2014ApJ...783..121G} also proposed three low-gravity L-type BDs as candidate members to this association. Since the probability for the Argus membership of SIMP~J2154-1055 is derived from a naive Bayesian classifier, it is expected to be biased when using dependent measurements as input observables (such as is the case here, see detail in \citealp{2014ApJ...783..121G}). Hence, we used the Monte Carlo contamination analysis presented in \cite{2014ApJ...783..121G} to obtain a probability of 20.5\% that SIMP~J2154--1055 is a young false-positive from the field. In the present case, the contamination and membership probabilities are almost exactly complementary, but this is not true in general. \\

\begin{figure*}
	\begin{center}
 	\includegraphics[width=0.99\textwidth]{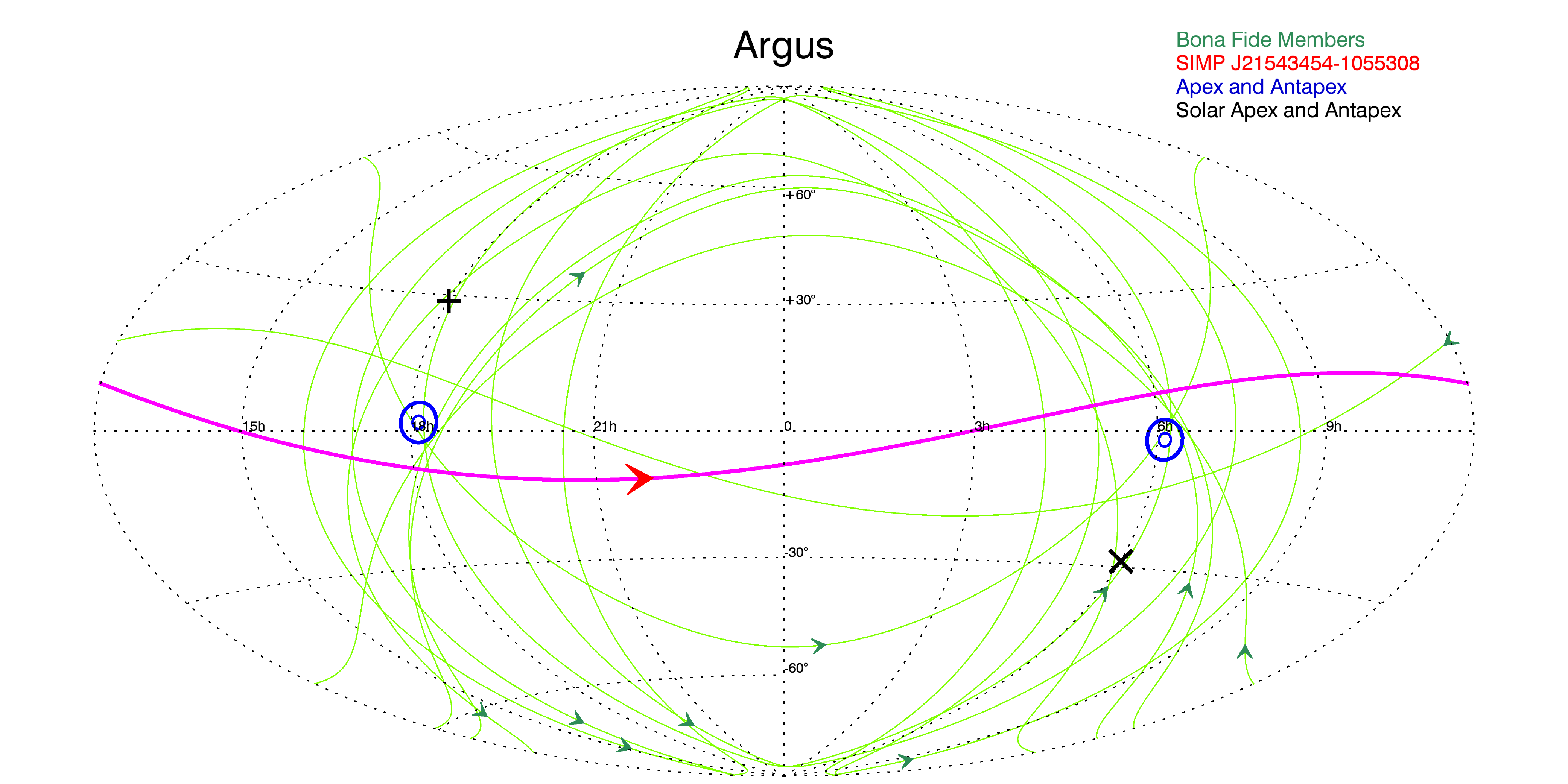}
	\end{center}
	\caption{Position and proper motion (red arrow) of SIMP~J2154--1055 with its associated projected great circle on the celestial sphere (pink line), as compared with bona fide members in the Argus association (green lines and arrows). It can be seen that the proper motion of SIMP~J2154--1055 is consistent with other bona fide members: its projected great circle falls closer to Argus' apex and antapex (blue circles) than 6 out of 11 known members. The solar motion's apex and antapex are displayed with a black plus sign and cross, respectively.}
	\label{fig:pm_graph}
\end{figure*}

In Figure~\ref{fig:pm_graph}, we show the proper motion of SIMP~J2154--1055 compared with those of known bona fide members in Argus. Since members are spread across a large portion of the sky, the direction of their proper motion can be different, however it is expected that the great circles corresponding to the motion of all members of a given moving group will pass close to the group's apex and antapex. The proper motion of SIMP~J2154--1055 thus seems consistent with a membership to Argus, since its great circle is closer to the apex than 6/11 bona fide members. Its very red $J-K_S$ colors (Figures~\ref{fig:Allers} and \ref{fig:2154}) are also consistent with this interpretation. The BANYAN~II tool allows us to predict that this object should have a distance of $22.1^{+2.8}_{-2.4}$~pc and a radial velocity of $-13.0~\pm~1.4$~\kms\ if it is a member of Argus, or a distance of $23.7^{+6.4}_{-5.2}$~pc and a radial velocity of $-8.9~\pm~9.1$~\kms\ if it is a field object. Using its 2MASS and ALLWISE apparent magnitudes, statistical distance, AMES-COND isochrones \citep{2003A&A...402..701B} and CIFIST2011 BT-SETTL atmosphere models \citep{2013MSAIS..24..128A} we determined that, at the age of Argus (30--50~Myr; \citealp{2008hsf2.book..757T}), SIMP~J2154--1055 has a predicted mass of 10--11~\Mjup. Measurements of radial velocity and parallax will be needed to assert its membership.\\

\section{CONCLUSIONS}\label{sec:conclusions}

We present the discovery of a new young L4$\beta$ BD, identified as part of the SIMP survey. We classify it as intermediate-gravity BD following the NIR gravity classification scheme of \cite{2013ApJ...772...79A}. We determine that it has an 84\% probability of being an Argus member. We estimate that if it is a member of the Argus association with an adopted age of 30--50~Myr, its mass would be 10~$\pm$~0.5~\Mjup, in the planetary-mass regime. Adding this object to the currently known late-type low-mass stars and BDs will help shape our understanding of the properties of low-gravity, low-pressure atmospheres reminiscent of those of giant, gaseous exoplanets. The data presented in this Letter can be downloaded at our group's Web site.\footnote[2]{\url{http://www.astro.umontreal.ca/\textasciitilde gagne}} \\

\acknowledgments

The authors would like to thank the anonymous referee for its suggestions, as well as Kelle Cruz and Jacqueline Faherty for useful comments and discussions and the sharing of spectroscopic templates. We thank Katelyn Allers for the sharing of several low-gravity spectra. This work was supported in part through grants from the Fonds de Recherche Qu\'eb\'ecois - Nature et Technologie and the Natural Science and Engineering Research Council of Canada. This research has benefited from the SpeX Prism Spectral Libraries, maintained by Adam Burgasser at \url{http://pono.ucsd.edu/\textasciitilde adam/browndwarfs/spexprism}, as well as the M, L, T and Y dwarf compendium housed at DwarfArchives.org. This research made use of; the SIMBAD database and VizieR catalogue access tool, operated at the Centre de Donn\'ees astronomiques de Strasbourg, France; data products from the 2MASS, which is a joint project of the University of Massachusetts and the Infrared Processing and Analysis Center (IPAC)/California Institute of Technology (Caltech), funded by the National Aeronautics and Space Administration (NASA) and the National Science Foundation \citep{2006AJ....131.1163S}; data products from the \emph{Wide-field Infrared Survey Explorer}, which is a joint project of the University of California, Los Angeles, and the Jet Propulsion Laboratory (JPL)/Caltech, funded by NASA \citep{2010AJ....140.1868W}; the NASA/IPAC Infrared Science Archive, which is operated by JPL, Caltech, under contract with NASA; the Infrared Telescope Facility (IRTF), which is operated by the University of Hawaii under Cooperative Agreement NNX-08AE38A with NASA, Science Mission Directorate, Planetary Astronomy Program; and of tools provided by Astrometry.net. This publication uses observations obtained at IRTF through program number 2008B054. The authors recognize and acknowledge the very significant cultural role and reverence that the summit of Mauna Kea has always had within the indigenous Hawaiian community. We are most fortunate to have the opportunity to conduct observations from this mountain.

%\bibliography{/Users/gagne/ApJ_Library}
%\mbox{~}

\end{document}